# Impact of Grain Boundaries on Efficiency and Stability of Organic-Inorganic Trihalide Perovskites

Zhaodong Chu[†1], Mengjin Yang[†2], Philip Schulz[2], Di Wu[1], Xin Ma[1], Edward Seifert[1], Liuyang Sun[1], Kai Zhu*[2], Xiaoqin Li*[1], Keji Lai*[1]

[1]Department of Physics, University of Texas at Austin, Austin, Texas 78712, USA

[2]National Renewable Energy Laboratory, Golden, Colorado 80401, USA

[†] These authors contributed equally to this work

* E-mails: kai.zhu@nrel.gov; elaineli@physics.utexas.edu; kejilai@physics.utexas.edu

# Abstract

Organic-inorganic perovskite solar cells have attracted tremendous attention because of their remarkably high power conversion efficiencies (PCEs). To further improve the device performance, however, it is imperative to obtain fundamental understandings on the photo-response and long-term stability down to the microscopic level. Here, we report the first quantitative nanoscale photoconductivity imaging on two methylammonium lead triiodide (MAPbI$_3$) thin films with different PCEs by light-stimulated microwave impedance microscopy. The intrinsic photo-response is largely uniform across grains and grain boundaries, which is direct evidence on the inherently benign nature of microstructures in the perovskite thin films. In contrast, the carrier mobility and lifetime are strongly affected by bulk properties such as the sample crystallinity. As visualized by the spatial evolution of local photoconductivity, the degradation due to water diffusion through the capping layer begins with the disintegration of large grains rather than the nucleation and propagation from grain boundaries. Our findings provide new insights to improve the electro-optical properties of MAPbI$_3$ thin films towards large-scale commercialization.

The worldwide surge of research interest in organic-inorganic trihalide perovskites, e.g., methylammonium lead triiodide ($CH_3NH_3PbI_3$ or $MAPbI_3$), has led to a phenomenal increase of the power conversion efficiency (PCE) of perovskite solar cells (PSCs) from 3.8% to 22% in the past few years.[1-7] These hybrid organic-inorganic thin films are polycrystalline in nature and compatible with low-cost solution or vapor-based processes.[8-10] Yet their performance rivals many single-crystalline semiconductor solar cells[11] owing to a number of intriguing optical and electrical properties that are ideal for energy harvesting and charge transport, such as high absorption coefficient across the solar spectrum,[12] high carrier mobility,[13,14] long carrier recombination lifetime.[15,16] The unprecedented progress of PSC efficiency was often attributed to the unique defect structures in the bulk and the benign grain boundaries during the early stage of PSC development.[17,18]

As the PSC efficiency continues to increase, recent efforts on increasing grain sizes and/or passivating grain boundaries (GBs) have started to cast doubts on the general belief in the unique defect tolerance in perovskites. Several groups including us have found that, when the grain size is increased from a few hundred nanometers to the micrometer level, the device performance is often significantly improved together with elongated charge carrier lifetimes[19-23]. At first sight, these studies imply that GBs in polycrystalline perovskite thin films may not be as benign as early studies had suggested. A recent theoretical study pointed out that GBs may even be the major recombination sites in the standard iodide based perovskites[24], which seems to be consistent with the recent experimental efforts described above. However, we should point out that the various new growth controls for increasing grain sizes could also affect the surface and bulk properties of perovskite grains, e.g., enhanced crystallinity, reduced defect density at the surface and in the bulk, and reduced structural defects associated with pinhole formation. Thus, it is important to scrutinize and isolate the impact of these different microscopic factors on electro-optical properties of polycrystalline perovskite thin films. Moreover, as material stability continues to be the key challenge faced by the PSC community[9], an immediate question is whether the GB and/or the surface of perovskite films are the weakest points where the degradation would start first. To this end, understanding the degradation mechanism at the microscopic level is also imperative for fabricating robust and reliable devices that meet the stringent requirements of commercialization[25-31]. In contrast to conventional macroscopic device characterizations, it is expected that spatially resolved studies on the chemical, electrical, and optical properties of the PSC thin films will provide crucial information for advancing the basic science and developing commercial products based on these fascinating materials.

While a number of scanning probe techniques[32-40] have been used to interrogate properties of the PSC, local measurements of the intrinsic photoconductivity, rather than the extrinsic photocurrent across the Schottky-like tip-sample junction, have not been reported to investigate the role of various microstructures on the films. In addition, due to the poor air stability of the organic-inorganic trihalide compound, experiments on samples with a surface encapsulation layer are usually desired to obtain reliable results. A non-contact method capable of mapping out the nanoscale photoconductivity on encapsulated films is thus particularly important for the PSC research. In this work, we report the first quantitative microwave impedance imaging with light stimulation on two $MAPbI_3$ thin films with different PCEs capped by a PMMA protection layer. The photo-response is spatially uniform across grains and grains boundaries, whereas the difference of carrier mobility and lifetime between these two films is attributed to the difference of sample crystallinity. Both the surface topography and local photoconductivity have been monitored over an extended period of time, which sheds new light on the intricate degradation process of the PSC devices.

Our experiments are performed on a unique microwave impedance microscopy (MIM)[41,42] setup with a focused laser beam illuminating from below the sample stage, as illustrated in Figure 1a. The bottom illumination scheme ensures an accurate calibration of the areal power intensity on the sample, which is not shadowed by the body of the cantilever probe[43]. Taking advantage of the near-field interaction, we are able to obtain a spatial resolution with deeply subwavelength lateral dimensions, i.e., on the order of the tip diameter (~ 100 nm or 1/30000 $\lambda$)[42]. Here the 1 GHz signal is delivered to the shielded cantilever tip[44] and the reflected microwave is amplified and demodulated to form MIM-Im and MIM-Re signals, which are proportional to the imaginary and real parts of the tip-sample admittance[41]. Unlike the conductive AFM,[34-37] a direct contact between the MIM tip and the perovskite thin films is not required due to the efficient capacitive coupling at GHz frequencies, overcoming the challenges of interrogating nanoscale electrical properties on samples with an insulating capping layer. The contrast mechanism of the MIM is similar to that of the time-resolved microwave conductivity (TRMC) experiment,[13,14,45] a non-contact technique widely exploited to study the dynamics of photo-generated carriers, except that the time resolution in TRMC under pulsed excitation is traded for spatial resolution in MIM under continuous illumination.

**Results**

**Microwave photoconductivity imaging on MAPbI$_3$.** The 100-nm thick MAPbI$_3$ thin films are prepared by an excess organic salt based solvent-solvent extraction method[23]. The precursor solution (30wt%) is spin-coated on glass substrates at 6000 rpm for 25 s to form a wet precursor film, which is then transferred into a stirred ether bath for 1 min, followed by thermal annealing at 150 °C for 15 min with a petri dish covered. Solar cell devices made from the same material but with thicker film (~350 nm) have demonstrated a PCE of 18% under the standard Air Mass (AM) 1.5 illumination (Supplementary Information, Figure S1). As a comparison, 100-nm thick MAPbI$_3$ films with smaller grain size was deposited using the same process except with stoichiometry precursor and annealing at 100 °C for 10 min. Devices with small grain size and thicker (~350 nm) perovskite film possess ~15% PCE (Figure S1). As plotted in Figure 1b, the two thin (~100 nm) films with different PCEs show very similar wavelength-dependent absorption and incident photon-to-electron conversion efficiency (IPCE, η) over the visible spectrum. Consistent with previous studies,[25-27] uncapped MAPbI$_3$ thin films degrade within several hours under the ambient light and humidity conditions, as shown in Figure S2. To prevent such rapid degradation, we have capped the sample surface with a thin (~30 nm thick) layer of spin-coated polymethyl methacrylate (PMMA)[46] in the following experiments. As shown in the schematic in Figure 1a, part of the PMMA/MAPbI$_3$ films is scratched away to expose the glass substrate as the reference region for the MIM imaging.

Figures 1c and 1d display the AFM/MIM images of the samples with 15% and 18% PCEs, respectively. The large difference in grain sizes between these two films is evident from the topographic data. Without the illumination, there is little MIM contrast between the highly resistive perovskite film and the glass substrate (Figure S3). When illuminated by an above-gap continuous-wave 532nm laser with a power intensity $P$ of 100 mW/cm$^2$ (equivalent to 1 Sun at AM 1.5), the 18% PCE sample exhibits clear photo-induced MIM signals, while the effect is much weaker on the 15% PCE sample. As a control experiment, we also performed the MIM imaging when the 18% PCE sample is illuminated by a below-gap 980-nm laser with the same power density of 100 mW/cm², as shown in Figure S3. The absence of photo-induced MIM signals with below-gap illumination further confirms the photoconductivity in the MAPbI$_3$ films.

**Quantitative analysis of photo-response.** The intrinsic photoconductivity images in Figure 1 contain much information on the photon-to-electron conversion process, which is at the heart of solar cell devices. For the quantitative analysis, we first focus on the power dependence of MIM response on the 18% PCE sample. Figure 2a shows the average MIM signals over an area of 3.5 μm × 3.5 μm as a function of the 532nm-laser intensity (raw data included in Figure S4). The MIM-Im signals increase monotonically with the laser power and saturate for both $P < 10$ mW/cm² and $P > 10^4$ mW/cm². The MIM-Re signals, on the other hand, reach a peak at $P \sim 10^3$ mW/cm² and decrease on both sides. Figure 2b shows the simulated MIM response as a function of the MAPbI$_3$ conductivity $\sigma$ using finite-element analysis[41] (FEA, detailed in Figure S5). The close resemblance between Figure 2a and 2b allows us to quantitatively extract $\sigma$ from the measured data. Since the film thickness is comparable to the tip diameter, the MIM is measuring the effective photoconductivity $\sigma$ averaged over the vertical direction, which can be derived as follows[42] (assuming complete absorption by the film).

$$\sigma = \frac{\eta}{H}\left(\frac{P\tau_e}{h\nu} \cdot e \cdot \mu_e + \frac{P\tau_h}{h\nu} \cdot e \cdot \mu_h\right) \quad (1)$$

As an order-of-magnitude estimate, i.e., taking the IPCE $\eta = 1$, film thickness $H = 100$ nm, electron and hole lifetime $\tau_e \approx \tau_h \approx 10^2$ ns, photon energy $h\nu = 2.33$ eV, electron and hole mobility $\mu_e \approx \mu_h \approx 10^2$ cm²/V·s, we obtain $\sigma \approx 10$ S/m at $P = 100$ mW/cm², which is in excellent agreement with the data in Figure 2c. The corresponding electron (hole) density is $n$ ($p$) = $P\tau_{e(h)}/h\nu \approx 3\times10^{15}$ cm⁻³, consistent with other investigations.[12] Note that the linear $\sigma \propto P$ relation extends down to $P \approx 1$ mW/cm². According to Eq. (1), our result implies that the carriers are not localized at a low density of $3\times10^{13}$ cm$^{-3}$, again consistent with studies reporting the exceptionally low trap density of $10^{10} \sim 10^{11}$ cm$^{-3}$ in MAPbI$_3$.[16] For laser intensity above 100 mW/cm$^2$, the sub-linear power dependence observed in Figure 2c may result from the saturation of photo-generated carriers.[48]

We now turn to the comparison between the two perovskite thin films. Following the same procedure, the power-dependent photoconductivity of the 15% PCE sample has been measured (raw data and FEA simulation shown in Figure S6) and plotted in Figure 2c. In the regime relevant to practical solar cell applications ($P \leq 100$ mW/cm$^2$), the photoconductivity of the 18% PCE sample is $5 \sim 6$ times higher than that of the 15% PCE sample. From the time-resolved photoluminescence (TRPL) data in Figure 2d, the average carrier lifetime differs by a factor of $\sim 3.5$ between the two films. As a result, the average mobility is also improved by a considerable amount ($\sim 1.5$ times) in the 18% film. Finally, from our X-ray diffraction (XRD) data in Figure 2e, the substantially better crystallinity of the PCE 18% film is vividly demonstrated by its much stronger PbI$_3$ peak than the 15% counterpart. In all, as the average grain size increases from a few hundred nanometers to a few micrometers due to the better growth control, both the electron-hole recombination rate and transport scattering rate are significantly reduced. The reduction of bulk and surface defect densities and the improvement of sample crystallinity are clearly important for future material and device optimization.

**Photoconductivity of grain boundaries.** To understand the nature of GBs during the photo-excitation, we directly compare the local photoconductivity between grains and GBs in the 18% PCE sample. From the AFM image in Figure 3a, appreciable inhomogeneity is seen on the polycrystalline MAPbI$_3$ thin film, with grain sizes ranging from sub-micrometer to a few micrometers and GBs appearing as trenches of $\sim 10$ nm in depth. In contrast, the corresponding

MIM images are quite uniform over grains with different heights and lateral sizes. In Figure 3b, we compare the line profiles across a single GB from the AFM and MIM images. Note that the MIM-Im channel is susceptible to the topographic crosstalk, while the MIM-Re signal is less sensitive to the surface roughness. For quantitative analysis of the GB photoconductivity $\sigma_{GB}$, we performed the 3D FEA simulation (detailed in Figure S7), in which the GB is modeled as a thin slab (width $w$ = 10 nm) with a height difference of $\Delta H$ = 10 nm lower than the grains ($\sigma_G$ = 10 S/m). Within the experimental uncertainty, a comparison between the MIM data and the FEA shows that $\sigma_{GB}$ is within a factor of two from $\sigma_G$. The result of nearly uniform photoconductivity is striking since GBs contain a large amount of dangling bonds, which usually lead to trap states detrimental to carrier generation and charge transport in conventional solar cell materials.[8] For halide perovskites, the role of GBs has been elusive over the years. While some theoretical studies showed that GBs only create shallow in-gap states due to the high ionicity and strong Pb-I anti-bonding,[17,18] others indicated that GBs are the major recombination centers and should be passivated.[24] Our measurements provide direct experimental evidence that, at least for encapsulated thin films with micrometer-sized grains, GBs are relatively benign to carrier generation and transport. Perovskite films with larger grains and better crystallinity, on the other hand, should be the direction to further improve the efficiency of PSCs.

**Spatial evolution of the degradation process**. The air stability of the perovskite films, which is arguably the most serious bottleneck towards commercial applications,[25-31] has also been studied by the MIM. Figure 4a shows selected AFM and MIM images of a PMMA-coated MAPbI$_3$ film over one week under the ambient condition with a relative humidity of 35% at 23°C (complete set of data included in Figure S8). Little change in the photoconductivity was observed on the first day or two, suggesting that the PMMA capping layer was initially effective in preventing the film degradation. Starting from Day 3, however, appreciable reduction of the photo-response was seen in certain regions of the film. Interestingly, the degradation does not emerge from GBs but is associated with the disintegration of large grains, which substantially reduces the photoconductivity in the surrounding regions. The degraded areas continued to grow during the subsequent days and the photoconductivity completely diminished after Day 6. As a control experiment, we also continuously illuminated another sample for 9 hours (same as the total imaging time in the aging experiment above) and observed no change in both topographic and photoconductivity maps (Figure S9). The illumination by itself, therefore, is not sufficient to cause the drastic suppression of photoconductivity in Figure 4a.

To gain further insight on the aging process, we carried out structural and optical characterizations on films prepared by the same method and exposed to the same conditions. As seen in the XRD results in Figure 4b, the characteristic $PbI_2$ peak at $2\theta = 12.66°$ appears on Day 1 and becomes significant on Day 3. The spectrum on Day 6 is dominated by the strong $PbI_2$ peak, indicative of the nearly complete decomposition of $MAPbI_3$ after one week. Note that the emergence of $PbI_2$ reduces the local photoconductivity because its absorption edge at 525nm[49] is below our 532nm laser. The time-resolved photoluminescence (TRPL) data plotted in Figure 4c also demonstrate the same trend. While little change is observed after one day, the absolute PL intensity drops substantially on Day 3 and the TRPL decay rate increases rapidly on Day 6. Interestingly, the TRPL decay rates are similar for the first three days, suggesting that the carrier lifetime remains large in the unaffected regions. Combining the microscopic and macroscopic results, we conclude that the deterioration of the encapsulated film is initiated and accelerated by the water molecules slowly diffused through the PMMA layer, which drive the decomposition of large MAPbI$_3$ grains into PbI$_2$ in the solid phase and CH$_3$NH$_2$ and HI in the gas phase.[25-28] Again, GBs are not the nucleation centers in the degradation process.

**Discussion**

To summarize, we report the intrinsic photoconductivity mapping of thin-film MAPbI$_3$ with different PCEs under above-gap illuminations. The large photoconductivity of the higher efficiency sample is a direct consequence of the high carrier mobility and long lifetime because of the improved crystallinity. Surprisingly, the grain boundaries exhibit photo-responses comparable to the grains, and they are not the nucleation centers for the degradation process. Our results highlight the unique defect tolerance responsible for the remarkable performance of perovskite solar cell devices, and address the significance of crystallinity to further improve their energy conversion efficiency. We also emphasize that the nanoscale photoconductivity imaging by microwave microscopy represents a new methodology in optoelectronic research. With future development to incorporate broadband illumination, variable temperatures, and humidity-controlled environment, we expect to obtain further insights on the intriguing photo physics and air sensitivity of the hybrid perovskite materials.

**Materials and Methods**

**Preparation of MAPbI$_3$ Perovskite Thin Films**. The perovskite films in this work were deposited on top of cover glasses using the stoichiometry and non-stoichiometry solvent-solvent extraction method.[23] In short, 30wt% methylammonium iodide (MAI) and lead iodide (PbI$_2$) (MAI/PbI$_2$=1/1, MAI/PbI$_2$=1.2/1) were dissolved in a mixed solvent of 1-N-methyl-2-pyrrolidinone (NMP) and $\gamma$-butyrolactone (GBL) (NMP/GBL=7/3 weight ratio). The cover glass with 80 $\mu$l precursor solution was spun at 6000 rpm for 25 s, and immediately dipped into a vigorously stirred diethyl ether bath for 1 min. The perovskite films rapidly crystallized during the bathing process and were further annealed (stoichiometry precursor: 100 ºC for 10 min; non-stoichiometry precursor: 150 °C for 15 min) with a petri-dish covered to remove excess organic salt. The encapsulated perovskite films were capped with polymethyl methacrylate (PMMA, $M_w$ ~120000) film by spin-coating 15 mg/ml PMMA in chlorobenzene solution at 4000 rpm for 35 s.

**Structural and Optical Characterizations**. The absorption spectra were measured by a UV/Vis spectrometer (Cary-6000i). The perovskite structure was characterized by an X-ray diffractometer (Rigaku D/Max 2200) using the Cu Kα radiation. The incident photon-to-electron conversion efficiency (IPCE) experiment was carried out in a solar cell quantum efficiency measurement system (QEX10, PV Measurements). The time resolved photoluminescence (TRPL) measurements were conducted in a home-built time correlated single photon counting system, where light source is a Fianium Supercontinuum high power broadband fiber filter (SC400-2-PP). The excitation wavelength was 500 nm with a spot area of 300 μm × 300 μm, a power of about 5 μW, and a repetition rate of 0.1 MHz.

**Microwave Impedance Microscopy**. The MIM in this work is based on a modified ParkAFM XE-100 platform with bottom illumination. The customized shielded cantilever probes[42] are commercially available from PrimeNano Inc. Finite-element analysis is performed using the AC/DC module of commercial software COMSOL4.4.

ASSOCIATED CONTENT

**Supporting Information**.

Figures S1 to S7. This material is available free of charge via the Internet at http://pubs.acs.org.

AUTHOR INFORMATION

**Corresponding Authors**

* E-mails: kai.zhu@nrel.gov; elaineli@physics.utexas.edu; kejilai@physics.utexas.edu

**Author Contributions**

K.L. X.L. and K.Z. conceived the project. M.Y., P.S., and K.Z. prepared samples and performed optical measurements. Z.C., X.M. and J.S. conducted the MIM measurements. Z.C. and K.L. performed data analysis and drafted the manuscript. The manuscript was written through contributions of all authors. All authors have given approval to the final version of the manuscript. Z.C. and M.Y. contributed equally to this work.

**Notes**

The authors declare no competing financial interest.

ACKNOWLEDGMENT

The work at UT-Austin led by X.L. and K.L. is supported by NSF EFMA-1542747. Z.C and D.W. also acknowledge the support from Welch Foundation Grant No. F-1814. The work at the National Renewable Energy Laboratory was supported by the U.S. Department of Energy under Contract No. DE-AC36-08-GO28308. K.Z., M.Y., and P.S. acknowledge the support by the hybrid perovskite solar cell program of the National Center for Photovoltaics funded by the U.S. Department of Energy, Office of Energy Efficiency and Renewable Energy, Solar Energy Technologies Office.

FIGURES

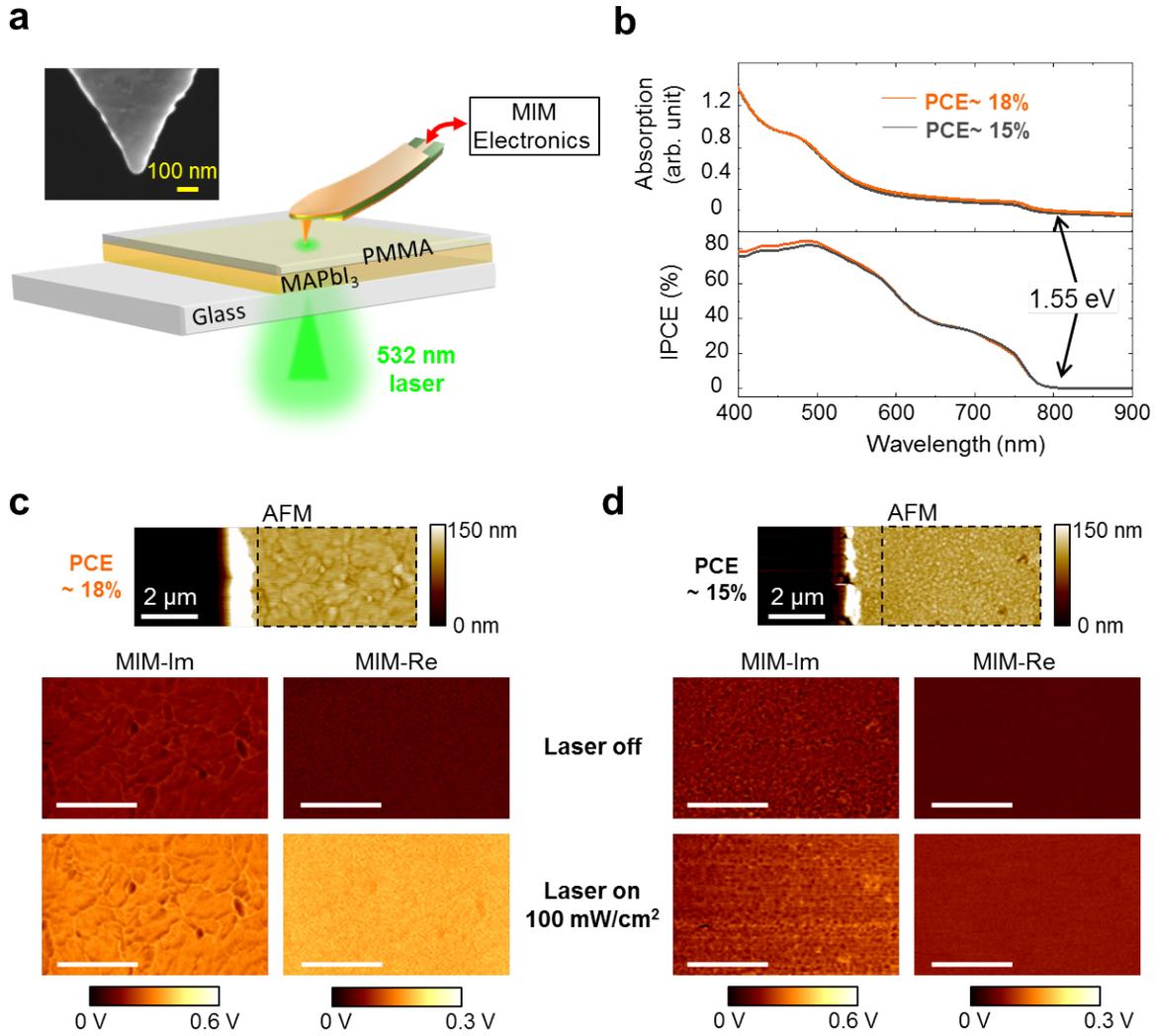

**Figure 1. Nanoscale photoconductivity imaging on MAPbI$_3$.** (a) Schematic diagram of the sample and the MIM setup with bottom illumination through the transparent glass substrate. Scanning is accomplished by moving the sample stage while fixing the laser beam and the probe tip, which are aligned before the experiment. The inset shows the scanning electron micrograph (SEM) of a typical MIM tip apex. (b) Optical absorption (upper panel) and IPCE (low panel) spectra of the perovskite thin films with 15% and 18% PCE. The optical gap of MAPbI$_3$ is indicated in the plots. (c) AFM (top) and MIM-Im/Re images (inside the black dashed rectangles

in the AFM data) when the 532-nm laser is turned off (middle) and on (bottom) for the 18% PCE sample. The same results for the 15% PCE sample are shown in (d). All scale bars are 2 μm.

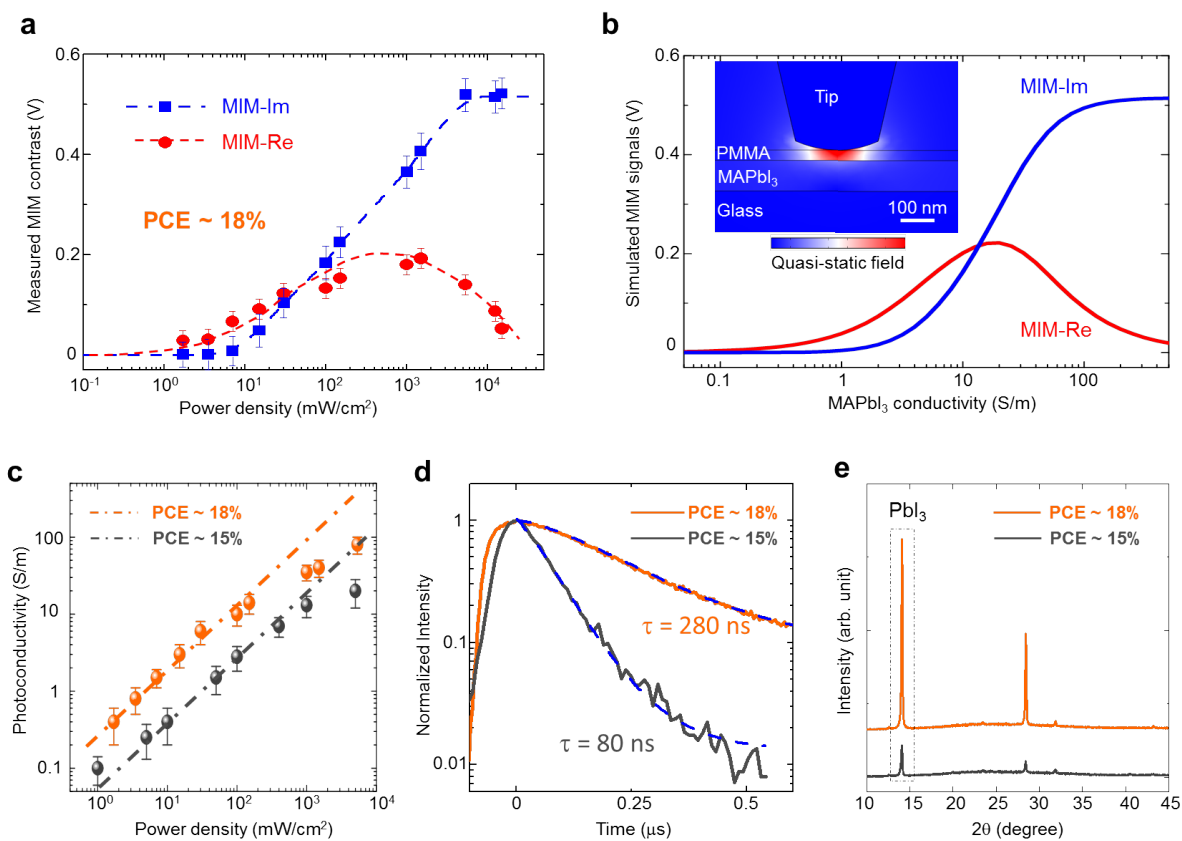

**Figure 2. Photoconductivity, carrier lifetime, and XRD of 18% and 15% PCE MAPbI$_3$ thin films**. (a) Measured MIM signals of the 18% PCE sample (averaged over an area of 3.5 μm × 3.5 μm, with respect to the values without light) as a function of the incident 532nm-laser power. The raw data are included in Fig. S3. (b) Simulated MIM response of the high-PCE sample as a function of the conductivity of the MAPbI$_3$ layer. The inset shows the simulated quasi-static displacement field distribution at the tip-sample junction. (c) Photoconductivity of both samples extracted from the experimental data versus the laser intensity. The dash-dotted lines are linear fits for laser power below 100 mW/cm$^2$. (d) TRPL measurements of the two samples, from which the carrier lifetime can be extracted. (e) XRD measurements of the two samples, indicating a much higher crystallinity of the 18% PCE sample.

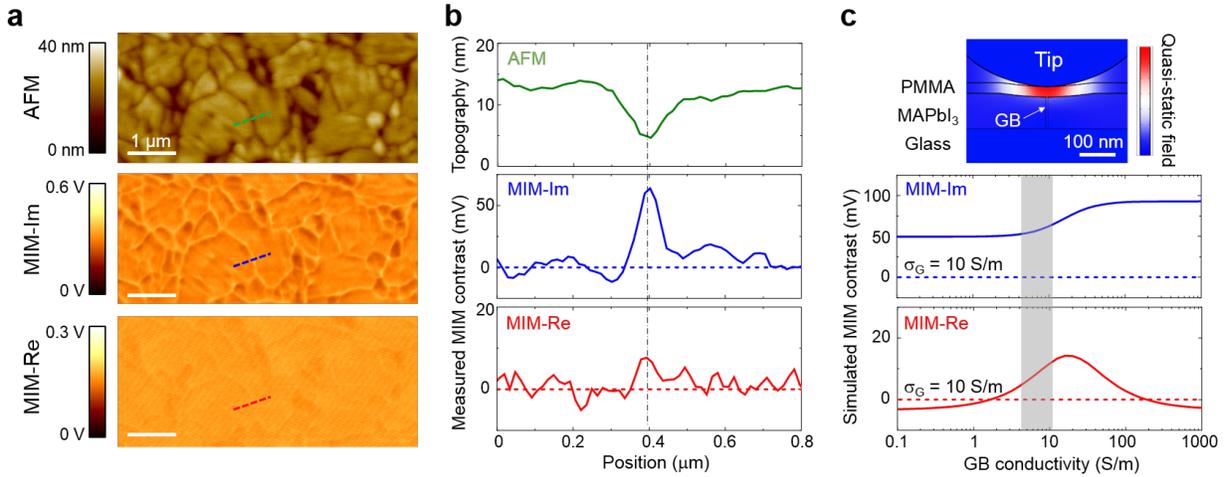

**Figure 3.** Various microstructures on the perovskite film. (a) AFM and MIM images showing the detailed features on the MAPbI$_3$ thin film under 100 mW/cm$^2$ illumination of the 532nm laser. The photo response is uniform for grains with different heights and lateral sizes. All scale bars are 1 μm. (b) Line profiles across a single grain boundary (green in AFM, blue in MIM-Im, and red in MIM-Re) in (a). The MIM contrast signals are referenced to the mean values on the grains. (c) Simulated MIM contrast, referenced to the background grain with a conductivity of 10 S/m, as a function of the GB conductivity $\sigma_{GB}$. The shaded column corresponds to the measured data within the experimental errors. The inset shows the simulated quasi-static displacement field distribution for $\sigma_G = \sigma_{GB}$ = 10 S/m.

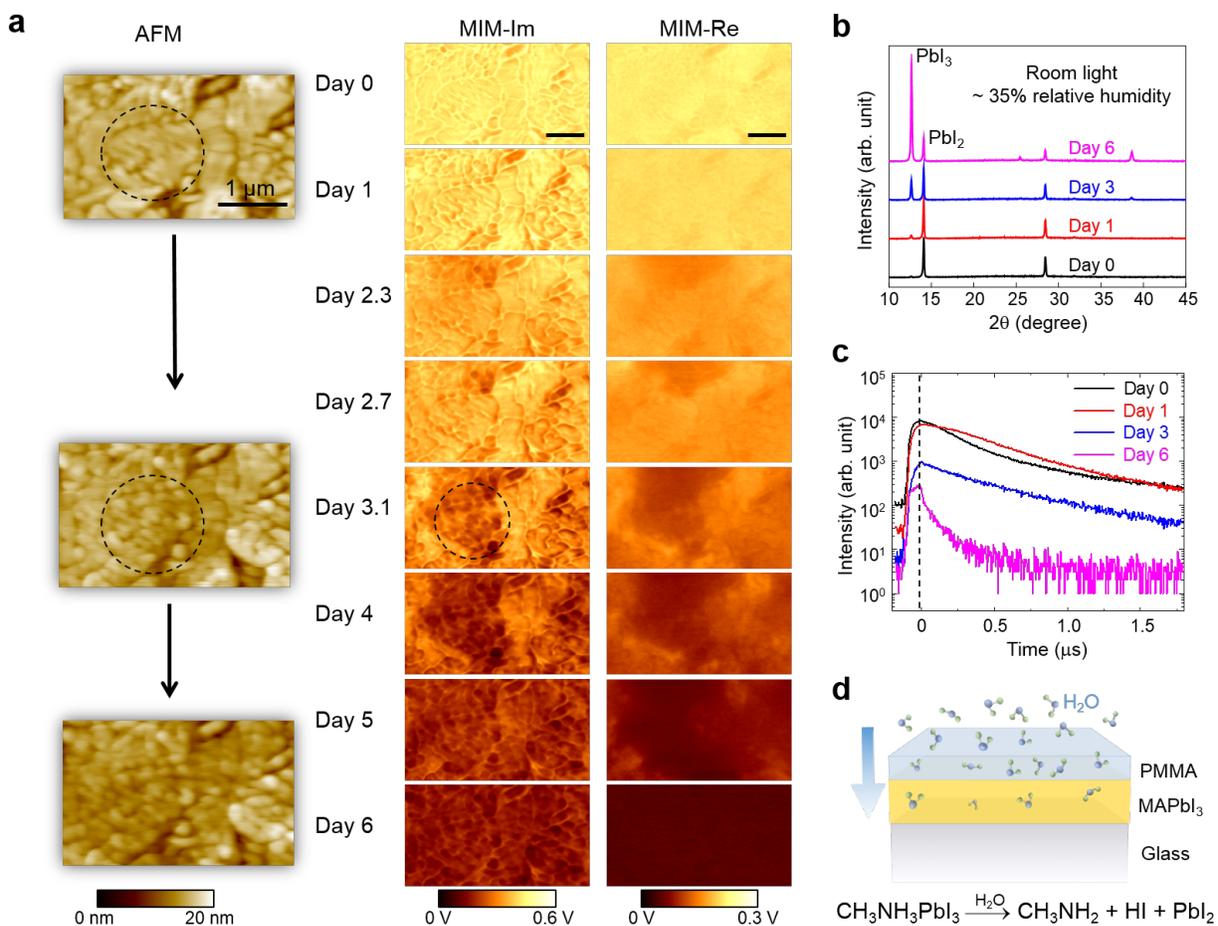

**Figure 4.** Degradation process of the MAPbI$_3$ thin film. (a) AFM and MIM images over one week with 35% relative humidity at 23°C. The sample was only illuminated by the 532-nm laser during the MIM imaging for about 30 min at each frame. A large grain inside the dashed circle on Day 0 was disintegrated into small grains after 3 days. All scale bars are 1 μm. (b) Selected XRD and (c) TRPL data on MAPbI$_3$ films prepared by the same method and exposed to the same conditions. The XRD peak at 2θ = 12.66° is associated with PbI$_2$. (d) Schematic of the aging process due to the diffusion of water molecules through the PMMA layer, which drives the decomposition of MAPbI$_3$ into CH$_3$NH$_2$ and HI in the gas phase, and PbI$_2$ in the solid phase.

23